\documentstyle[aps,prb,psfig,multicol]{revtex}
\begin{document}
\draft
\title{Raman scattering in a two-dimensional electron gas: Boltzmann
equation approach}
\author{E. G. Mishchenko\footnote{Present address: Lorentz-Instituut, Leiden University, P.O. Box 9506, 
2300 RA Leiden, The Netherlands}}

\address{L.D. Landau Institute for Theoretical Physics,\\
Russian Academy of Sciences, Kosygin 2, Moscow 117334, Russia}
\maketitle

\begin{abstract}
The inelastic light scattering in a 2-d electron gas is
studied theoretically
using the Boltzmann equation techniques.
Electron-hole excitations produce the Raman spectrum essentially
different from the one predicted for the 3-d case.  
In the clean limit it
has the form of a strong non-symmetric resonance
due to the square root singularity at the electron-hole 
frequency $\omega = vk$ while
in the opposite dirty limit the usual Lorentzian shape of
the cross section is reestablished.
The effects of electromagnetic field are considered self-consistently
and the contribution from collective plasmon modes is found.
It is shown that
unlike 3-d metals where plasmon excitations
are unobservable (because of very large required transfered
frequencies), the two-dimensional electron
system gives rise to a low-frequency
($\omega \propto k^{1/2}$) plasmon peak. A measurement of the width of this
peak can provide data on the magnitude of the electron scattering rate.

\end{abstract}

\begin{multicols}{2}

\par Raman scattering is a powerful method for experimental
studies of elementary excitations in various structures.
In particular, high-$T_c$ superconductors produce
Raman spectra which remain mysterious over a broad region
of frequency. Namely, the high-frequency continuum \cite{SKR,RKL},
$2\Delta$-peak (see, e.g., Ref. \cite{BKK}), two-magnon spectra, revealing strong mutual
influence between antiferromagnetism and superconductivity \cite{RRD},
still does not have a robust self-consistent theoretical description.
Therefore, the development of the theory of Raman scattering from
different excitations is still of considerable current interest.
\par Here we are interested in the Raman scattering from excitations
of a 2-d normal electron system, namely from electron-hole pairs and collective
plasmon excitations. We show that  Raman scattering cross
section in  2-d systems
differs from the spectrum of a  3-d metal in two aspects.
First, the scattering from electron-hole pairs becomes more singular 
(due to the square-root singularity in the density of states). A
 finite strength of the electron-hole contribution is determined by the 
electron
scattering rate.
Second, as soon as the plasmon spectrum is gapless 
($\omega \propto k^{1/2}$), a corresponding peak is located in 
the reasonably low-frequency ($\sim 10~ meV$) range.
\par Inelastic light scattering in 2-d systems has been
extensively used in investigations of the excitations 
of the Fermi sea \cite{ACP}, the energy gap in the fractional
quantum Hall regime \cite{PVH}, exciton states \cite{ZPC},
spin-density and charge-density excitations \cite{DPV}.
Experimental evidence of 2-d plasmons in GaAs heterostructures
comes from the Raman spectroscopy measurements in a magnetic field \cite{KKS}.
By varying the direction of magnetic field it is possible to distinguish
Raman response of a 2-d electron system from a contribution of a background 
\cite{BMP}.
The $k^{1/2}$-spectrum is clearly observed\cite {KKS}, however, no data are 
available for
the dependence of a lineshape of the plasmon peak on the electron scattering
rate.
\par The standard quantum mechanical theory of Raman scattering in electron
systems 
applies the 
Green function formalism. We use in the present paper a different
approach, based on the Boltzmann equation. Such a semiclassical method
is valid \cite{LL} when the characteristic scale of transferred light momentum 
is less than the Fermi momentum of a 2-d electron gas. For the typical
situation of GaAs/AlGaAs heterojunction the concentration of carriers
\cite{PVH,KKS}
is of the order of $1 - 6 \times 10^{11} cm^{-2}$ and corresponds
to the Fermi momentum value of $\sim 3 - 8 \times 10^5 cm^{-1}$. On the
other hand the typical
values of the momentum transfer are considerably lower
 $\sim 0.2 - 1 \times 10^{5} cm^{-1}$. The main advantage of the kinetic approach is
a possibility to include  effects 
of electron scattering and to account for electromagnetic field
in a self-consistent manner by solving simultaneously 
 the Maxwell equation. This
gives a plasmon contribution together with the electron-hole continuum.
It is also possible to  generalise easily  this theory to the case of 
external magnetic field applied. 
\begin{figure}[h]
  \unitlength 1cm
  \begin{center}
  \begin{picture}(8,5)
 \put(-3.5,-7.2){\includegraphics{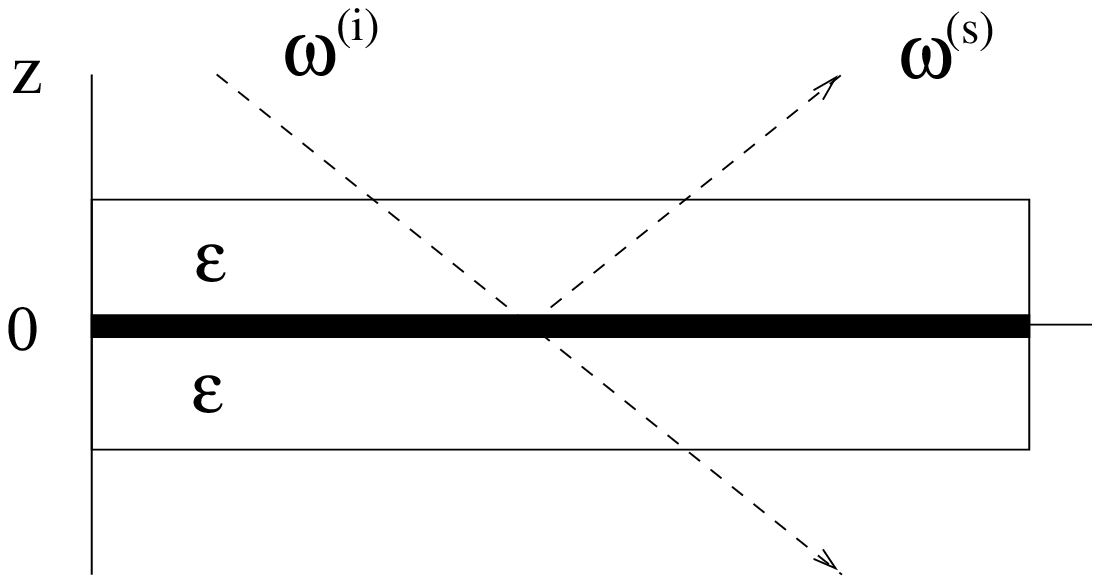}}
  \end{picture}
  \end{center}
  \caption[]{
 Geometry of the light scattering.
Thick solid line ($z=0$ plane) represents the electron gas
embedded into a host material with the dielectric constant $\epsilon$.
The incident (i) light induces the in-plane 2-d current which in its turn
produce a scattered (s) wave with different frequency.}
\label{fig1}
\end{figure}
\par
The system under study is shown in Fig. 1.
Two-dimensional  electron gas ($z=0$ plane) is embedded into a host
material with the dielectric constant $\epsilon$.
The polarization of incident (i) and, hence, of scattered (s) light
waves are assumed to be parallel to the plane $z=0$.
The effective Hamiltonian describing Raman scattering from
electronic fluctuations is bilinear in the vector potential
of light,

\begin{equation}
\label{ham}
H_{eff}=\frac{e^{2}}{2mc^2}\int d^2s ~\delta n_{\gamma}({\bf s},t)
{\bf A}^{2}({\bf s},t),
\end{equation}
where the fluctuation $\delta n_{\gamma}$ expressed via the nonequilibrium
partition
function $\delta f_p({\bf s},t)$
\begin{equation}
\label{par}
\label{flu}
\delta n_{\gamma}({\bf s},t)=\int \frac{2d^2 p}{(2\pi)^2} \gamma_p
\delta f_p({\bf s},t),
\end{equation}
differs from the usual electronic density only by the
anisotropic dimensionless factor $\gamma_p$ (electron-light vertex).
This factor depends on the light polarization and accounts for the virtual 
interband transitions. Its exact form is not essential 
for the following (see Ref. \cite{AG}).
\par Varying the expression (\ref{ham}) over the vector potential ${\bf A}$
we obtain the electron current induced by the incident
light with the frequency $\omega^{(i)}$ and the in-plane
wave vector ${\bf k}_{s}^{(i)}$:
\begin{equation}
\label{cur}
{\bf j}^{(i)}({\bf s},t)= -\frac{e^2}{mc} \delta n_{\gamma}
({\bf s},t)
{\bf A}^{(i)}\exp{(-i\omega^{(i)}t+i{\bf k}_{s}^{(i)}{\bf s})}.
\end{equation}
The 2-d current (\ref{cur}) produces a scattered electromagnetic wave with
different frequency  $\omega^{(s)}$ and wave vector ${\bf k}_{s}^{(s)}$.
The solution of corresponding nonuniform Maxwell equation is straightforward.
After some simple calculations it gives for the amplitude of light scattered into
 the half-space $z>0$ the expression,
\begin{equation}
A(\omega^{(s)},{\bf k}_{s}^{(s)}) =\frac{2\pi ie^2  A^{(i)}}
{mc^2 k_{z}^{(s)}} \delta n_{\gamma} (\omega, {\bf k}_{s}),
\end{equation}
where $k_z^{(s)2}=\epsilon \omega^{(s)2}/c^2 -{\bf k}_{s}^{(s)2}$; 
the Fourier-component of the density fluctuations depends on
the transferred energy $\omega =\omega^{(i)} -\omega^{(s)}$ and
momentum ${\bf k}_{s}= {\bf k}_{s}^{(i)} -
{\bf k}_{s}^{(s)}$.
\par The Raman scattering  cross section defined as the normalised energy
flow   $\left< \left< \vert \omega^{(s)} A ^{(s)}/ \omega^{(i)}
A ^{(i)}\vert^{2} \right> \right>$
related to the interval $d\omega^{(s)}d^2k_{s}^{(s)}/(2\pi)^3$,
has the form:
\begin{equation}
\label{sec}
\frac{d^2 \sigma}{d\omega^{(s)} do^{(s)}}=
\frac{\epsilon^{1/2} e^4}{2\pi m^2 c^5}\frac{\omega^{(s)3}}{\omega^{(i)2}
k_z^{(s)}} K(\omega,{\bf k}_{s}),
\end{equation}
where $K(\omega,{\bf k}_{s})$ is the Fourier component of the correlator
of density 
fluctuations 
$$
K({\bf s} -{\bf s'},t-t')=
\left< \left< \delta n_{\gamma}({\bf s},t) \delta n_{\gamma}({\bf s'},t')
\right> \right>.
$$
One can argue that the expression (\ref{sec}) diverges as the
direction of scattered light approaches  the electron plane:
 $k_z^{(s)} \rightarrow 0$.
In fact, this means that as soon as the "width" of a two-dimensional system $l$
is assumed to be the smallest one of all of the characteristic lengths of the
problem, we are restricted to the limit $k_z^{(s)} >> l^{-1}$.

To evaluate this correlator we apply the fluctuation-dissipation
theorem which expresses  it via the imaginary part of the
generalized response $\delta n_{\gamma}  (\omega, {\bf k})$ to an arbitrary external
potential $U(\omega, {\bf k})$ (in what follows we omit the subscript $s$):

\begin{equation}
\label{fdt}
K(\omega, {\bf k}) =-\frac{2}{1-\exp{(-\omega/T)}}~
\mbox{Im} \left(\frac{\delta n_{\gamma} (\omega, {\bf k})}{
U (\omega, {\bf k})} \right).
\end{equation}

The  most simple way to derive the generalised response (\ref{fdt})
is to make use of the linearized
Boltzmann equation for the nonequilibrium part of the
distribution function
$\delta f_p =\chi_p
\partial f_0/\partial \varepsilon$ \cite{FK,FM}:
\begin{equation}
\label{bol}
-i(\omega -{\bf kv} + i\tau^{-1}) \chi_p({\bf k},\omega)=
i\omega \gamma_p U ({\bf k},\omega) - e {\bf vE}({\bf k},\omega),
\end{equation}
where $f_0(\varepsilon)$ is the local-equilibrium Fermi-Dirac partition function.
The second term in the right hand side of Eq. (\ref{bol}) accounts for the fluctuating
electromagnetic field  ${\bf E}$. It satisfies the Maxwell
equation with the nonequilibrium electric current determined from Eq. (\ref{bol}):
\begin{equation}
\label{max}
\mbox{rot rot} ~ {\bf E}(z,{\bf s},\omega)
-\frac{\epsilon\omega^2}{c^2} {\bf E}(z,{\bf s},\omega)
= -\frac{4\pi i e\omega}{c^2}\delta(z)
\left< {\bf v} \chi_p({\bf s},\omega)\right>.
\end{equation}
Here the brackets denote the integral over the Fermi line
$$
\left< ... \right> = \int \frac{2dp_F}{v (2\pi)^2}(...).
$$
\par We are interested in the solution of Eq. (\ref{max})
at $z=0$. The straightforward derivation gives for the
Fourier component of the electric field 
\begin{equation}
\label{sol}
{\bf E}(z=0,{\bf k},\omega) = \frac{2\pi i e}{\epsilon \omega}
\sqrt{k^2- \epsilon \omega^2/c^2}
\left< {\bf v} \chi_p({\bf k},\omega)\right>.
\end{equation}
Substituting the solution (\ref{sol}) into the Boltzmann equation
(\ref{bol}) one gets the integral equation for the electronic density fluctuation
$\chi_p({\bf k},\omega)$. Such an equation has a simple solution 
 which after the substitution into
Eq. (\ref{flu}) and then into
the fluctuation-dissipation theorem (\ref{fdt}) gives
the Raman cross section.
Finally we obtain (see Fig. 2)
\begin{eqnarray}
\label{cor}
K ({\bf k},\omega) \propto -\mbox{Im} ~
\left< \frac{\omega\gamma_p^2}{\omega -{\bf kv} +i\tau^{-1}}
\right> + \nonumber \\
 \mbox{Im} ~ F_{\alpha}({\bf k},\omega) D_{\alpha \beta}
({\bf k},\omega)F_{\beta}({\bf k},\omega),
\end{eqnarray}
where the proportionality coefficient
(Bose factor) is omitted,
see Eq. (\ref{fdt});
$D({\bf k},\omega)$ is the two-dimensional electromagnetic
Green function
\begin{equation}
\label{gre}
D^{-1}_{\alpha \beta}({\bf k},\omega)=
\frac{1}{\omega} \left< \frac{ v_{\alpha} v_{\beta}}{\omega - {\bf kv} +i\tau^{-1}} \right> -
\frac{\epsilon \delta_{\alpha \beta}}{2\pi e^2 \sqrt{k^2 -\epsilon \omega^2/c^2}}
\end{equation}
and $F_{\alpha}({\bf k},\omega)$ is the oscillator strength
\begin{equation}
\label{str}
F_{\alpha}({\bf k},\omega)=
\left< \frac{ v_{\alpha}\gamma_p}{\omega -
{\bf kv} +i\tau^{-1}} \right>.
\end{equation}
We devote the rest of the paper to the discussion of different
terms in the Raman cross
section (\ref{cor}).
\begin{figure}[h]
  \unitlength 1cm
  \begin{center}
  \begin{picture}(8,2)
 \put(0,0.5){\includegraphics{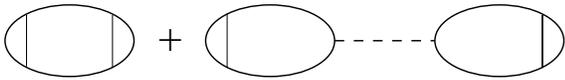}}
  \end{picture}
  \end{center}
  \caption[]{Diagrammatic representation of the 
density-density correlation function (\ref{cor}). The solid lines 
represent electron propagators, the dashed line stands for the
electromagnetic Green function and the empty vertex for the effective 
electron-light interaction.}
\label{fig2}
\end{figure}

\par The first term represents the scattering from the
electron-hole pairs. For the estimates we suppose  the  Fermi line
to be
isotropic, i.e.,
\begin{equation}
\label{e-h}
K_{e-h}(k,\omega) \propto
-\frac{m \gamma^2}{\pi}
 ~ \mbox{Im} ~
\frac{\omega}{\sqrt{ (\omega+i\tau^{-1})^2 - k^2v^2}}.
\end{equation}
\par In the dirty limit $kv\tau <<1$ (the so-called zero-momentum transfer
limit)  one can neglect $v^2k^2$
in the denominator. The cross section then takes the same
well-known Lorentzian form as in 3-d case \cite{ZC}:
$\sim \omega\tau/ (\omega^2\tau^2 +1)$.
\par In the clean limit $kv\tau >>1$ the expression (\ref{e-h})
has
a square root singularity at $\omega=kv$ rather
than a step-like one (as in 3-d case).
It results in the strong non-symmetric resonance (see Fig. 3);
the finite  height of this resonance is controlled by the scattering
rate $\tau^{-1}$.
For anisotropic Fermi line the resonance location is
defined by the maximum value of electron velocity
along the momentum transfer $\omega = {\bf kv}_{max}$.
\par The second term in Eq. (\ref{cor})
represents effects of Coulomb interaction
and collective electron excitations, namely 2-d plasmons.
This term is important in the clean limit only.
At low transferred frequencies $\omega << kv$ it results
in 
the screening of the isotropic scattering channel through
the renormalization of the electron-light
vertex $\gamma_p \rightarrow
\gamma_p - \left< \gamma_p \right>/\left<1\right>$,
similar to usual 3-d case \cite{FK,FM}.
\par At high transferred frequencies $\omega >> kv$
the second term in Eq. (\ref{cor}) gives a 2-d plasmon
peak located at the plasmon frequency 
$\omega_{pl}(k)$ which  is determined
by the dispersion equation
\begin{equation}
\label{pla}
\omega^2=\frac{2\pi e^2}{\epsilon} \left< v_{k}^2 \right>
\sqrt{k^2 -\epsilon \omega^2/c^2},
\end{equation}
where $v_k$ means the component of electron velocity 
in the $k$-direction.
The typical momenta transfer $k$ are of the order of
light momenta. Hence from the formula
(\ref{pla}) one can see that $\omega >> vk$ proving 
the initial assumption was correct. We can also omit 
the term $\epsilon\omega^2/c^2$ (this term cares for
the finite plasmon velocity) in Eq. (\ref{pla})
in comparison with the term $k^2$ due to the fact that
$c^2k>>v^2 p_F$ for typical values of $k$.
Indeed, this means that it is enough to use the Poisson
equation for electromagnetic fluctuations instead of the  Maxwell
equation (\ref{max}). The only difference occurs at very 
small transferred
momenta where
the Poisson equation gives the infinite plasmon velocity
in the limit $k \rightarrow 0$. 
\par The formula (\ref{pla})
is valid only if the plasmon wavelength becomes large compared to the
layer thickness, $k>>l^{-1}$. If this condition is violated the
three-dimensional problem has to be solved with boundary conditions
satisfied on both sides of the layer. It's solution gives the expression
$$
\omega^2_{pl}(k)=\frac{4\pi e^2}{\epsilon} 
\sqrt{\left< v_{k}^2 \right> \left< v_{z}^2 \right>}
~ \mbox{tanh}\left( \sqrt{\frac{\left< v_{k}^2 \right>}{
 \left< v_{z}^2 \right>}} \frac{kl}{2}
\right),
$$
where  the electron velocity along the
perpendicular direction $v_{z}$ appeares. Note, that the angular brackets
now denote the integral over the three-dimensional Fermi surface.
When $l\rightarrow 0$ this expression reduces to Eq. (\ref{pla}) and
for $l\rightarrow \infty$ it gives the frequency of ordinary 3-d plasmon.
\begin{figure}[h]
  \unitlength 1cm
  \begin{center}
  \begin{picture}(8,5)
 \put(0.5,0){\includegraphics{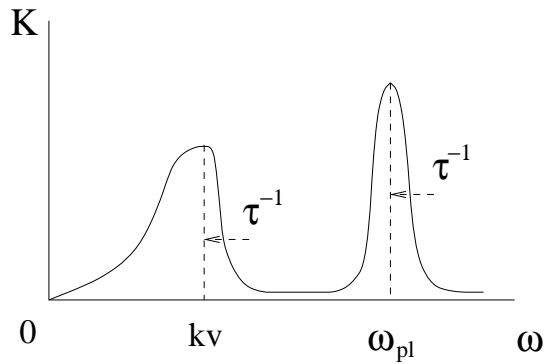}}
  \end{picture}
  \end{center}
  \caption[]{ Raman scattering cross section.
The non-symmetric resonance is located at $\omega =kv$.
Its width at $\omega > kv$ as well as its height are determined
by the finite value of the scattering rate $\tau^{-1}$.
The second symmetric resonance corresponds to excitation
of plasmon.}
\label{fig3}
\end{figure}

\par Near the plasmon resonance the Raman cross section has the 
symmetric Lorentzian lineshape 
\begin{equation}
\label{pea}
K_{pl} (k, \omega) \propto
\frac{m \gamma^2}{8\pi\omega}
\frac{k^2 v^2 \tau^{-1} }{
(\omega_{pl}(k)-\omega)^2+\tau^{-2}/4}.
\end{equation}
The relative height of two resonances (\ref{e-h}) and (\ref{pea})
is $K_{e-h}/K_{pl} \sim k^{3/2}v^{3/2}\tau^{1/2}/\omega_{pl}$
and can be either more or less than unity depending on the
momentum transfer $k$ and scattering rate $\tau^{-1}$.

\par In conclusion, we have calculated the Raman scattering 
intensity from two-dimensional
electronic fluctuations. The main distinctive features from
a usual three-dimensional metal are: the more singular electron-hole
contribution (\ref{e-h}) and low frequency plasmon resonance
(\ref{pea}).
The electronic Raman scattering
in a 2-d system in a transverse magnetic field  can be considered
studied  the same Boltzmann equation technique as it has been done for 3-d electron
system \cite{M}.
\par Author thanks Prof. L.A. Falkovsky for numerous fruitful discussions
and valuable comments. The work was supported by the Russian Foundation
for Basic Research, Grant No 97-02-16044 and by a scholarship from KFA, Forschungszentrum in Juelich, Germany.

\end{multicols}

\end{document}